\documentclass[runningheads]{llncs}
\usepackage[T1]{fontenc}
\usepackage{graphicx}
\usepackage{amsmath}
\usepackage{subcaption}  
\usepackage{url}
\usepackage{graphicx}
\usepackage{xcolor}

\begin{document}
%
%\title{Enhancing FCD Lesion Segmentation with Total Variation Regularization}

\title{A Total Variation Regularized Framework for Epilepsy-Related MRI Image Segmentation }

\titlerunning{Total Variation for 3D Epilepsy MRI Segmentation}
\author{Mehdi Rabiee\inst{1}\orcidID{0009-0004-0308-8033} 
\and
Sergio Greco\inst{1}\orcidID{0000-0003-2966-348} \and
Reza Shahbazian\inst{1}\orcidID{0000-0002-2313-6002} \and
Irina Trubitsyna\inst{1,2}\orcidID{0000-0002-9031-0672} 
 }
\authorrunning{M. Rabiee et al.}

\institute{
\inst{} Department of Computer Engineering, Modeling, Electronics and Systems (DIMES), University of Calabria, Rende 87036, Italy. \\
\and
\inst{}Corresponding author: i.trubitsyna@dimes.unical.it
}

\maketitle              % typeset the header of the contribution
\begin{abstract}
%Three-dimensional (3D) medical image segmentation is a challenging subject in computer vision due to high-dimensional input data, limited labeled samples, and the need for anatomically consistent outputs.  Focal Cortical Dysplasia (FCD) is a major cause of drug-resistant epilepsy, where lesions are subtle, small, and difficult to detect even for trained specialists. 
{Focal Cortical Dysplasia} (FCD) is a primary cause of drug-resistant epilepsy and is difficult to detect in brain {magnetic resonance imaging} (MRI) due to the subtle and small-scale nature of its lesions. Accurate segmentation of FCD regions in 3D multimodal brain MRI images is essential for effective surgical planning and treatment. However, this task remains highly challenging due to the limited availability of annotated FCD datasets,  the extremely small size and weak contrast of FCD lesions, the complexity of handling 3D multimodal inputs, and  the need for output smoothness and anatomical consistency, which is often not addressed by standard voxel-wise loss functions. This paper presents a new framework for segmenting FCD regions in 3D  brain MRI images. We adopt state-of-the-art transformer-enhanced encoder-decoder architecture and introduce a novel loss function combining Dice loss with an anisotropic {Total Variation} (TV) term. This integration encourages spatial smoothness and reduces false positive clusters without relying on post-processing. The framework is evaluated on a public FCD dataset with 85 epilepsy patients and demonstrates superior segmentation accuracy and consistency compared to standard loss formulations. The model with the proposed TV loss shows an 11.9\% improvement on the Dice coefficient and 13.3\% higher precision over the baseline model. Moreover, the number of false positive clusters is reduced by 61.6\%.

\end{abstract}
\section{Introduction}
%NOT connected to the nextparagraph. to deleet/move: Recent advances in data engineering have made it easier to integrate large amounts of medical imaging data into analytical frameworks.  Three-dimensional (3D) medical image segmentation tasks present significant problems due to high-dimensional input data, limited labelled samples, and the need for anatomically consistent outputs~\cite{zhong2025pmfsnet}. 

Epilepsy is a neurological disorder characterized by a persistent predisposition to generate unprovoked seizures, affecting millions worldwide and necessitating accurate diagnosis and management due to its potential long-term impact on quality of life and brain function~\cite{fisher2014ilae}. The 75th World Health Assembly and World Health Organization (WHO) selected epilepsy as one of the top priorities in the prevention and control of noncommunicable diseases~\cite{feigin2025global}.

Epilepsy is often linked to lesions or abnormalities on the brain's cortex, which trigger and spread seizures. The most common cause is focal cortical dysplasia (FCD), which encompasses a spectrum of developmental malformations in the cerebral cortex, marked by localized disruptions in cortical architecture and cellular composition~\cite{splitkova2025new}. It is considered the leading cause of drug-resistant epilepsy in children and remains a significant factor in the use of anti-epileptic medications among adults~\cite{jimenez2023automatic}. 
%FCD can be treated with medication or surgery to remove the affected brain area. 
Identifying FCD regions in brain magnetic resonance imaging (MRI) images is vital for successful surgery and a better chance of curing epilepsy.

Artificial intelligence (AI), and in particular deep learning (DL), can help find potential epilepsy regions.  The goal is to perform a medical image segmentation task where the inputs are 3D volumes consisting of voxels (similar to pixels in 2D images), usually reconstructed from a sequence of 2D MRI images recorded by medical imaging devices with known position and orientation of the recording device for each frame. 3D images usually have multiple modalities, which are captured with different parameters of the MRI scanner and can get different aspects of brain structure; for example, T1-weighted, T2-weighted, FLAIR, and PET~\cite{symms2004review}. Although the problem is similar to other medical imaging diagnosis tasks, like detecting brain tumors from MRI images or analyzing 3D medical images from CT scans, detecting FCD regions is more challenging because of their very small sizes that are hard to see even by experts. 
Another important issue is the availability of robust and annotated training data. Unlike many other tasks, there are only a few small-sized datasets available for FCD detection. Therefore, it is essential to utilize a robust architecture with optimal hyperparameters and training strategies to achieve the best possible results.

Medical image segmentation is a well-studied subject.
%there are a lot of studies and proposed network architectures in literature for medical image segmentation that usually can be applied to 2d or 3d images, however almost all of them perform better on 2d inputs rather than 3d. 
%
The state of the art for medical image segmentation is based on \textit{U-Net} architecture~\cite{ronneberger2015u}. This architecture is shaped like a letter \textit{U} that consists
of a symmetric encoder-decoder architecture with a contracting path for feature
extraction and an expansive path for precise localization.
%and consists of an encoder path in the down-going part of the \textit{U} (see Figure \ref{fig:arch}). The encoder essentially consists of convolution blocks, increasing  the number of channels by a factor of 2 at each level while simultaneously decreasing the image size by a factor of 2.  %to get a representation feature at the bottleneck, and a 
%As shown in Figure~\ref{fig:arch}, the decoder path in the upgoing part of the \textit{U} acts as the opposite of the encoder to construct the output prediction map from the feature. The image resolution is doubled at each level while the number of channels is decreased by a factor of~$2$.
A skip connection between the decoder and the corresponding encoder block at each level  enables the model to use fine detail information from encoders in the reconstruction path. This design allows \textit{U-Net} to efficiently learn spatial hierarchies and retain high-resolution contextual information, making it especially effective for pixel-wise segmentation tasks in biomedical imaging.
%There is a final output block which is a convolution layer that changes the number of layers to the expected number of output channels.
%The success of \textit{U-Net} inspired many similar architectures for medical image segmentation specially 
With the emergence of transformer architectures and their success in language processing and computer vision tasks, some studies combined the idea of transformers with the well-established \textit{U-Net} architecture.  In particular, vision transformers or attention blocks are used to capture long-range dependencies and global context, complementing convolutional features. Some well-performing architectures are \textit{UNETR}~\cite{hatamizadeh2022unetr}, \textit{Swin UNETR}~\cite{hatamizadeh2021swin}, \textit{UNETR++}~\cite{shaker2024unetr++}, and \textit{MS-DSA-Net} that outperforms the other existing methods in FCD detection task~\cite{zhang2024focal}.

\paragraph{Contributions. }
 In this paper we focus on a real-world clinical challenge: segmenting FCD regions in 3D brain MRI images,  
 and adopt the state-of-the-art method  \textit{MS-DSA-Net}~\cite{zhang2024focal} as the base. 
Due to the limited size and complexity of available FCD datasets, we carefully design a training pipeline based on patch-wise sampling and voxel-wise classification, enabling the model to learn effectively from limited and high-dimensional data.
%Instead of relying on post-processing to clean up noisy or fragmented segmentation outputs, we integrate spatial consistency directly into the training process. 
We propose a new loss by adding a Total Variation (TV) regularization term to the loss function, which encourages the model to produce smoother and more anatomically consistent segmentation masks by penalizing abrupt changes in neighboring voxel predictions.
We validate our proposed approach on a publicly available dataset of annotated FCD scans. In particular, we consider different combinations of Dice Loss, Cross Entropy Loss and Total Variation component in the presence and absence of post-processing that cleans up noisy or fragmented segmentation outputs. The results show that adding the TV regularization to a standard Dice loss not only improves segmentation accuracy but also leads to cleaner, more coherent prediction maps. This can improve the further advanced AI-based automated detections.

\paragraph{Organization. }
The rest of this paper is organized as follows:
Section~\ref{sec:2} reviews related works on deep learning models for medical image segmentation, including transformer-based architectures. Section~\ref{sec:3} introduces our proposed model, including the incorporation of Total Variation (TV) loss into the MS-DSA-Net architecture. Section~\ref{sec:4} describes the experimental setup, dataset, training details, and evaluation metrics. This section also presents quantitative and qualitative results, followed by an in-depth discussion. Finally, Section~\ref{sec:5} concludes the paper and outlines future research directions.

\section{Related Works}\label{sec:2}
%With the rapid growth of convolutional neural networks and their success in the field of computer vision and general image processing tasks, many researchers started to adapt the concepts and architectures to the more specialized domain of medical image processing which has some limitations compared to general image processing like fewer available datasets due to privacy and the high cost of labeling by experts, high dimensionality images and class imbalance of data and variation of different medical imaging devices. 
%U-Net [3]:
In this section we briefly describe the main architectures of 3D medical image segmentation.
U-Net~\cite{ronneberger2015u} was introduced in 2015 and since then has been utilized as the base architecture for state-of-the-art methods. The U-Net consists of a symmetric encoder-decoder architecture with a contracting path for feature extraction and an expansive path for precise localization. The contracting path applies repeated $3\times3$ convolutions (without padding), each followed by ReLU and $2\times2$ max pooling with stride 2, doubling the number of feature channels at each step. The expansive path upsamples the feature maps using $2\times2$ deconvolutions that halve the feature channels, concatenates them with the corresponding cropped feature maps from the encoder, and applies two $3\times3$ convolutions followed by ReLU. A final channel-wise convolution maps the output to the desired number of classes. 
%The U-Net architecture plays a role as a base for many other research projects, and it has been adapted by other researchers to use multimodal 3d image inputs as well.
%SegResNet [8]:
SegResNet~\cite{myronenko20183d} uses an encoder-decoder convolutional neural network (CNN) architecture with an asymmetrically larger encoder for feature extraction and a smaller decoder for mask reconstruction. To enhance training on limited data, a variational autoencoder (VAE) branch is added at the encoder’s endpoint to reconstruct the input image, providing additional regularization and guidance. The encoder is based on ResNet blocks using $3\times3\times3$ convolutions with \textit{Group Normalization} and ReLU, combined with strided convolutions for downsampling and skip connections for feature preservation. The encoder reduces spatial dimensions while increasing feature depth. The decoder mirrors this structure but uses fewer blocks per level, upsampling features via non-trainable 3D bilinear interpolation and combining them with corresponding encoder outputs. A final $1\times1\times1$ convolution and sigmoid function produce the segmentation output. The VAE branch compresses the encoder output into a latent representation (mean and standard deviation), samples from it, and reconstructs the image using a decoder-like path without skip connections.
The main features of SegResNet are using a VAE branch for better learning on small dataset sizes and using more blocks with residual connections in the encoder path compared to the decoder path and also using non-trainable operations in the upsampling process. 

After the introduction of transformers and attention mechanisms and their success in language modeling and consequently using them in image processing tasks like vision transformers (ViTs), some researchers utilized them in medical image segmentation tasks.
%UNETR [4]:
UNETR~\cite{hatamizadeh2022unetr} follows a U-Net-like encoder-decoder structure, where the encoder is built entirely from transformer blocks operating on a sequence of non-overlapping 3D image patches. The input volume is divided into uniform patches, flattened, and linearly projected into a fixed $K$-dimensional embedding space, with positional embeddings added to retain spatial context. The transformer encoder comprises multiple layers of multi-head self-attention (MSA) and MLP blocks, using residual connections and layer normalization. Feature maps are extracted from different transformer layers (layers 3, 6, 9, 12), reshaped back into 3D tensors, and connected to the decoder through skip connections. The decoder progressively upsamples the features using deconvolutional layers, combines them with corresponding encoder outputs via concatenation, and applies $3\times3\times3$ convolutions and normalization. A final $1\times1\times1$ convolution with softmax activation produces the voxel-wise segmentation output.
%Swin UNETR [5]:
Swin UNETR~\cite{hatamizadeh2021swin} builds on the UNETR architecture by replacing the standard transformer encoder with a Swin Transformer~\cite{liu2021swin}, which introduces a more efficient way to model self-attention in 3D medical images. While UNETR processes the entire 3D volume as a sequence of fixed-size patches and applies global self-attention across all patches, Swin UNETR computes self-attention within local windows and shifts these windows between layers to allow communication  between neighboring regions. This shifted window mechanism helps reduce computational cost while still capturing long-range dependencies.
%UNETR++ [6]:
UNETR++~\cite{shaker2024unetr++} builds on the UNETR architecture by introducing an efficient paired-attention (EPA) block to enhance feature representation. The EPA block combines spatial and channel attention using shared query-key pairs, enabling the model to efficiently capture both global spatial relationships and inter-channel dependencies. This dual-attention mechanism improves segmentation accuracy while maintaining low computational cost. Similar to the original U-Net, UNETR++ progressively reduces spatial resolution and increases the number of feature channels at each encoder stage.

%MS-DSA-Net [7]:
MS-DSA-Net~\cite{zhang2024focal} {also follows similar design principles to those used in the U-Net architecture~\cite{zhang2024focal}, which remains foundational for medical image segmentation tasks. Its peculiarity is the integration of the parallel transformer pathways with dual self-attention (DSA) modules to enhance lesion segmentation.}
 Each DSA module combines spatial and channel self-attention using shared queries and keys but separate value paths, capturing both inter-position and inter-channel dependencies efficiently. Features are fused in the decoder through deconvolution and fusion blocks to recover spatial detail and generate precise probability maps.  The reported results in~\cite{zhang2024focal} indicate that MS-DSA-Net shows the best performance among the existing architectures for the FCD detection task. Therefore, we adopt this architecture to apply the proposed TV loss.

\section{Proposed Model}\label{sec:3}
\begin{figure}
    \centering
    \includegraphics[width=\linewidth]{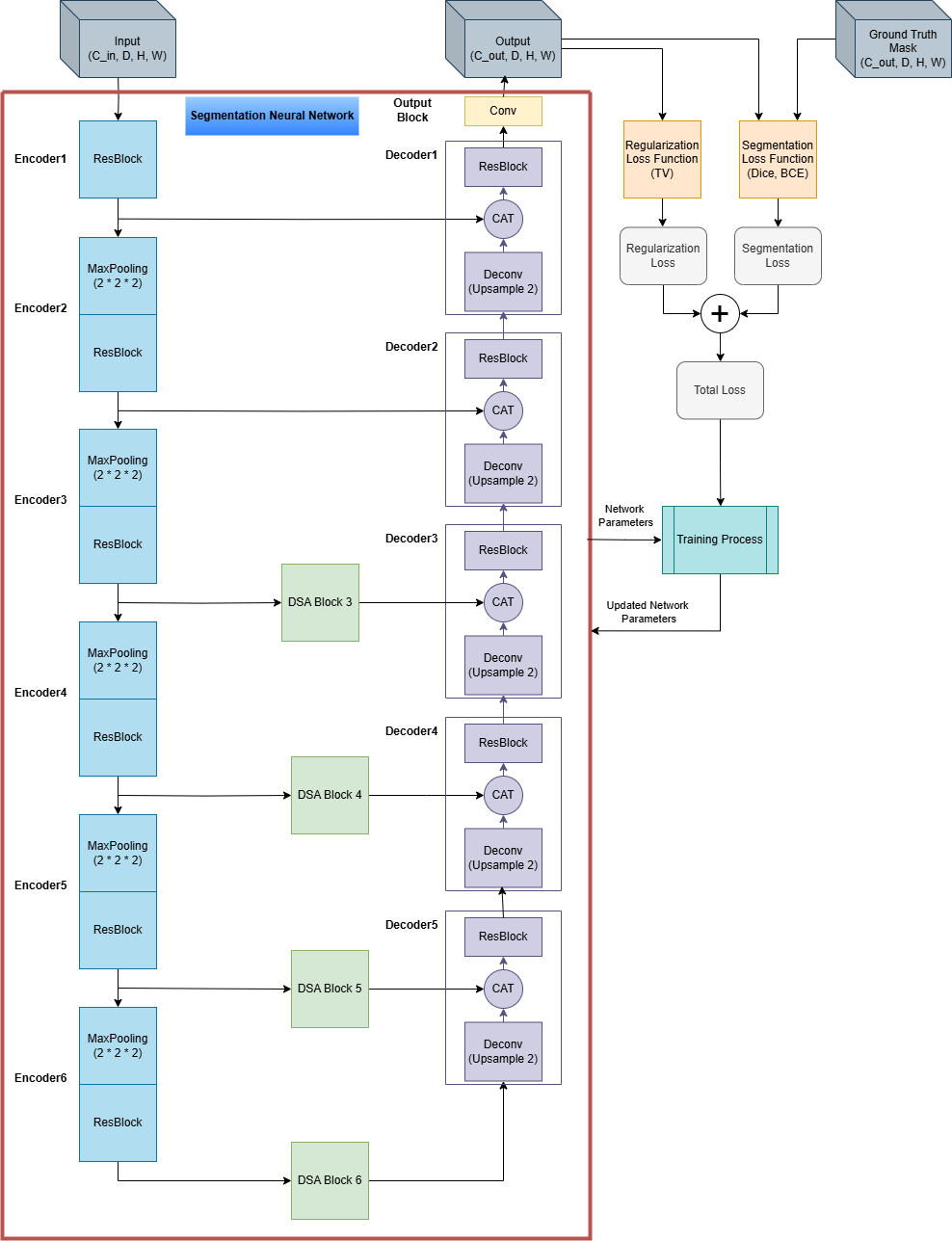}
    \caption{The architecture of the system for FCD detection. It is based on the \mbox{MS-DSA-Net} (inside the red box) and the proposed TV loss function. }
    \label{fig:arch}
\end{figure}

System architecture based on the MS-DSA-Net is given in Figure \ref{fig:arch}. 
%The most common loss functions that have been used in network training for medical image segmentation are 

The majority of studies, including the MS-DSA-Net, utilize \textit{Dice Loss}, \textit{Cross Entropy Loss,} or a combination of these two as the training loss. These loss functions are based on independent voxel prediction regarding the ground truth label of voxels. 
The main idea is to integrate a regularization term for output spatial smoothness in the loss function. This can teach the network to generate more consistent output maps so that the nearby voxels have similar probability values. To this end, we introduce an anisotropic Total Variation (TV) loss function. 

The Total Variation loss is defined over the predicted voxel values to encourage spatial smoothness in the segmentation output. 
%{The overall encoder-decoder structure, use of skip connections, and residual units follow similar design principles to those used in the U-Net architecture~\cite{zhang2024focal}, which remains foundational for medical image segmentation tasks.}

%For a predicted probability volume $p$, 
Let $p_{i,j,k}$ be the predicted probability at voxel location $(i,j,k)$, 
the isotropic TV loss $\mathcal{L}_{\text{TV}}$ {is formally defined as}:

\begin{equation}
\mathcal{L}_{\text{TV}} = \sum_{i,j,k} \left( |p_{i+1,j,k} - p_{i,j,k}| + |p_{i,j+1,k} - p_{i,j,k}| + |p_{i,j,k+1} - p_{i,j,k}| \right)
\end{equation}

%\begin{equation}
%\mathcal{L}_{\text{TV}} = \sum_{i,j,k} \left( 
%left| x p_{i,j,k} \right| +
%left| y p_{i,j,k} \right| +
%left| z p_{i,j,k} \right| \right)
%

%where:

%\begin{align*}
%x p_{i,j,k} &= p_{i+1,j,k} - p_{i,j,k} \\
%y p_{i,j,k} &= p_{i,j+1,k} - p_{i,j,k} \\
%z p_{i,j,k} &= p_{i,j,k+1} - p_{i,j,k} \\
%\noindent where $p_{i,j,k}$ is predicted probability at voxel location $(i,j,k)$.
%\end{align*}

TV loss is widely used in image denoising and super-resolution models due to its ability to suppress noise while preserving edges. To the best of our knowledge, no previous study has incorporated TV loss into a UNet-based architecture for volumetric medical image segmentation, such as the detection of FCD in 3D MRI. It should be mentioned that the idea has been introduced for some 2D image segmentation tasks, such as the study performed by Javanmardi et al.~\cite{javanmardi2016unsupervised}. 
In this paper, we integrate this regularization directly into the training loss function to promote contiguous and anatomically plausible segmentation in 3D space.

We adopt MS-DSA-Net~\cite{zhang2024focal} as the base architecture, as it achieves superior results on the FCD segmentation task among state-of-the-art methods. Our network consists of six encoder blocks, starting with 16 channels and doubling the number of channels at each stage until reaching 512 channels at the bottleneck. Correspondingly, the spatial resolution is halved at each stage via $2\times2$ max-pooling. Each encoder block includes a residual unit composed of two convolutional layers with instance normalization and leaky ReLU activation. The output of the two convolutions is added to the input through a residual connection, followed by another convolution and normalization layer.

The decoder path consists of five decoder blocks, each performing the inverse operations of its corresponding encoder: halving the number of channels and doubling the spatial resolution. Each decoder block starts with a deconvolution layer, then concatenates its output with the skip connection from the corresponding encoder (when available), and passes the result through a residual block similar to the encoder design.
The final output layer is a convolutional block that produces a prediction map with the same spatial size as the input but with two channels: one for the background and one for FCD.
Skip connections from the encoder to the decoder begin from stage 3 and employ dual self-attention transformers. These blocks consist of a channel-wise attention module and a spatial attention module with dimensionality reduction via a linear layer. The outputs of both attention modules are added to the input using residual connections. We used an input patch size of $128\times128\times128$, randomly selected from training subjects. 

\subsection{Loss Functions}
The base loss for training was Dice Loss computed only on the FCD channel. To evaluate the effect of integrating the proposed regularization, we utilize three loss formulations, explained as follows:

\paragraph{Dice Loss}\cite{zhang2024focal}:
\begin{equation}
\mathcal{L}_{\text{Dice}} = 1 - \frac{2 \sum_i p_i g_i + \epsilon}{\sum_i p_i + \sum_i g_i + \epsilon}
\end{equation}

\paragraph{Binary Cross Entropy (BCE) Loss:}
\begin{equation}
\mathcal{L}_{\text{BCE}} = -\frac{1}{N} \sum_{i=1}^{N} \left[ g_i \log(p_i) + (1 - g_i) \log(1 - p_i) \right]
\end{equation}
where $p_i$ is the predicted probability, $g_i$ is the ground truth label for voxel $i$, and $\epsilon = 1\times10^{-5}$ ensures numerical stability.

%\paragraph{TV Loss:}
%We define the Total Variation loss $\mathcal{L}_{\text{TV}}$ as follow:
%\begin{equation}
%\mathcal{L}_{\text{TV}} = \sum_{i,j,k} \left( |p_{i+1,j,k} - p_{i,j,k}| + |p_{i,j+1,k} - p_{i,j,k}| + |p_{i,j,k+1} - p_{i,j,k}| \right)
%\end{equation}

\paragraph{Total Loss:}
\begin{itemize}
  \item \text{Dice + BCE (equal weight):}
  \begin{equation}
  \mathcal{L}_{\text{Total}} = 0.5 \cdot \mathcal{L}_{\text{Dice}} + 0.5 \cdot \mathcal{L}_{\text{BCE}}
  \end{equation}
  \item \text{Dice + TV (TV weight = 0.1):}
  \begin{equation}
  \mathcal{L}_{\text{Total}} = 1.0 \cdot \mathcal{L}_{\text{Dice}} + 0.1 \cdot \mathcal{L}_{\text{TV}}
  \end{equation}
\end{itemize}

The weight of 0.1 is chosen for the TV loss because higher values may encourage trivial zero outputs, while lower values reduce its regularization impact (based on practical results). The TV loss is computed across three directions $(x, y, z)$ without averaging, making the effective regularization equivalent to weighting the average by 0.3.
{The coefficient 1.0 for the Dice loss ensures that it remains the primary component guiding the segmentation. The TV term is weighted at 0.1 to act as a regularizer. Although the total sum exceeds 1.0, the loss terms are on different scales, and this combination was selected based on empirical performance. A lower weight for the TV term prevents over-smoothing or trivial solutions (e.g., empty masks), while still encouraging spatial consistency.}

\section{Experiments} \label{sec:4}
\subsection{Dataset and Preprocessing}
We use the same dataset as in the MS-DSA-Net~\cite{zhang2024focal}. The dataset is available publicly~\cite{schuch2023open}. It consists of T1 and FLAIR MRI modalities from 85 epilepsy patients and 85 healthy controls. For this study, only patient data was used and split randomly into training, validation, and test sets. Preprocessing (reorientation, skull stripping, modality alignment, and registration to MNI152 template space) was performed using the FSL toolkit\footnote{https://fsl.fmrib.ox.ac.uk/fsl/docs/}.

\subsection{Training}
The training procedure involves the following steps:
\begin{itemize}
  \item Random patch sampling with balanced FCD/background samples
  \item Data augmentation including random crop, rotation, flipping, intensity shift, and adding Gaussian noise
  \item Input patches fed as batches into the network
  \item Loss computation and backpropagation using AdamW optimizer
  \item Early stopping based on validation loss stagnation (patience threshold)
  %\item {Early stopping was employed by monitoring the validation loss. Training was halted if the loss did not improve for a specified number of consecutive epochs (a patience threshold of 25).}

\end{itemize}
Weights were initialized using Kaiming normal for convolutional layers, Xavier uniform for linear and attention layers, and constants for normalization layers. A learning rate scheduler was used, beginning at 10\% of the maximum rate, linearly warming up for 10 epochs, and followed by cosine annealing decay. {Early stopping was employed by monitoring the validation loss. Training was halted if the loss did not improve for a specified number of consecutive epochs (a patience threshold of 25).} All experiments were implemented using PyTorch and MONAI, incorporating code from the MS-DSA-Net repository\footnote{https://github.com/zhangxd0530/MS-DSA-Net}.
We used the following configurations for training and evaluation:
\begin{itemize}
    \item \textbf{Subject Split:} 59 training, 12 validation, and 14 test subjects
    \item \textbf{Patches per Image:} 4
    \item \textbf{Initial Learning Rate:} $1\times10^{-4}$
    \item \textbf{Minimum Learning Rate:} $1\times10^{-6}$
    \item \textbf{Max Epochs:} 300
    \item \textbf{Batch Size:} 1 (i.e., 4 patches of one subject per batch)
    \item \textbf{Early Stopping Patience:} 25 epochs
    \item \textbf{Total Trainable Parameters:} 43,524,802
    \item \textbf{Hardware:} NVIDIA GeForce RTX 2080 Ti (12 GB RAM)
\end{itemize}

\subsection{Evaluation Metrics}
Voxel-level validation metrics after each epoch included sensitivity (Sens), precision (Prec), and mean Dice score (DC). On the test set, we also computed i) subject-level sensitivity (sSens): presence of any true positive voxel match
and ii) False Positive Clusters (nFPC): average number of falsely detected voxel clusters per subject.

\subsection{Post-processing}
After prediction, we applied connected component analysis with the following steps:
\begin{itemize}
  \item Binary opening: dilation followed by erosion
  \item Binary hole filling with $5\times5\times5$ kernel
  \item Connected component labeling with 26-connectivity ($3\times3\times3$ structure)
  \item Cluster size filtering: removal of clusters smaller than 50 voxels
\end{itemize}

These post-processing steps and subject-level metrics are inspired by the base method, although specific hyperparameters were not disclosed.
To account for randomness, each experiment was performed 10 times, and the reported values include the mean and standard deviation of the evaluated metrics. The same train, validation, and test splits were used across all scenarios.

\subsection{Quantitative Results}
As shown in Table~\ref{tab:results}, adding TV loss to the Dice loss leads to a noticeable improvement in the Dice score, both with and without post-processing. While the addition of BCE loss also shows gains, its impact is smaller compared to TV loss. 
Regarding the average number of false positive clusters (nFPC), TV loss consistently reduces this metric more effectively than BCE loss. Although BCE slightly improves voxel- and subject-level sensitivity more than TV, the precision is better when TV loss is used.
Comparing pre- and post-processed results, it is evident that post-processing improves Dice score, precision, and nFPC across all loss functions, albeit with a slight reduction in sensitivity. Notably, the gain from post-processing with plain Dice loss (from 0.2811 to 0.2866) is smaller than the gain achieved by adding TV loss to Dice (from 0.2811 to 0.3104). Additionally, the original nFPC value with Dice+TV is already significantly better than with Dice alone, indicating stronger inherent regularization.

\begin{table}[h!]
\centering
\caption{Evaluation results (mean $\pm$ standard deviation) on the test and post-processed test datasets: sensitivity (Sens), precision (Prec), mean Dice score (DC), subject-level sensitivity (sSens), and False Positive Clusters (nFPC).}
\begin{subtable}[t]{\linewidth}
\centering
\caption{Test Results}
\resizebox{\linewidth}{!}{%
\begin{tabular}{|l|c|c|c|c|c|}
\hline
\textbf{Loss} & \textbf{Sens} & \textbf{Prec} & \textbf{DC} & \textbf{sSens} & \textbf{nFPC} \\
\hline
Dice & 0.3822 ± 0.0337 & 0.2767 ± 0.0417 & 0.2811 ± 0.0200 & 0.7857 ± 0.0337 & 22.0071 ± 7.5851 \\
Dice + BCE & 0.3964 ± 0.0439 & 0.2648 ± 0.0824 & 0.2885 ± 0.0473 & 0.8071 ± 0.0345 & 9.8071 ± 6.4381 \\
Dice + TV & 0.3845 ± 0.0425 & 0.3000 ± 0.0428 & 0.3104 ± 0.0246 & 0.8000 ± 0.0301 & 8.4500 ± 2.2591 \\
\hline
\end{tabular}%
}
\end{subtable}

\vspace{-0.2cm}

\begin{subtable}[t]{\linewidth}
\centering
\caption{Post-processed Test Results}
\resizebox{\linewidth}{!}{%
\begin{tabular}{|l|c|c|c|c|c|}
\hline
\textbf{Loss} & \textbf{Sens} & \textbf{Prec} & \textbf{DC} & \textbf{sSens} & \textbf{nFPC} \\
\hline
Dice & 0.3765 ± 0.0333 & 0.2880 ± 0.0440 & 0.2866 ± 0.0204 & 0.7714 ± 0.0452 & 3.8786 ± 1.0118 \\
Dice + BCE & 0.3916 ± 0.0437 & 0.2735 ± 0.0858 & 0.2925 ± 0.0471 & 0.8071 ± 0.0345 & 3.1714 ± 1.0834 \\
Dice + TV & 0.3788 ± 0.0426 & 0.3102 ± 0.0444 & 0.3146 ± 0.0246 & 0.7928 ± 0.0226 & 3.1643 ± 0.7439 \\
\hline
\end{tabular}%
}
\end{subtable}
\label{tab:results}
\end{table}

\subsection{Qualitative Results}
\begin{figure}[h!]
    \centering
    \begin{subfigure}[t]{0.48\linewidth}
        \centering
        \includegraphics[width=\linewidth]{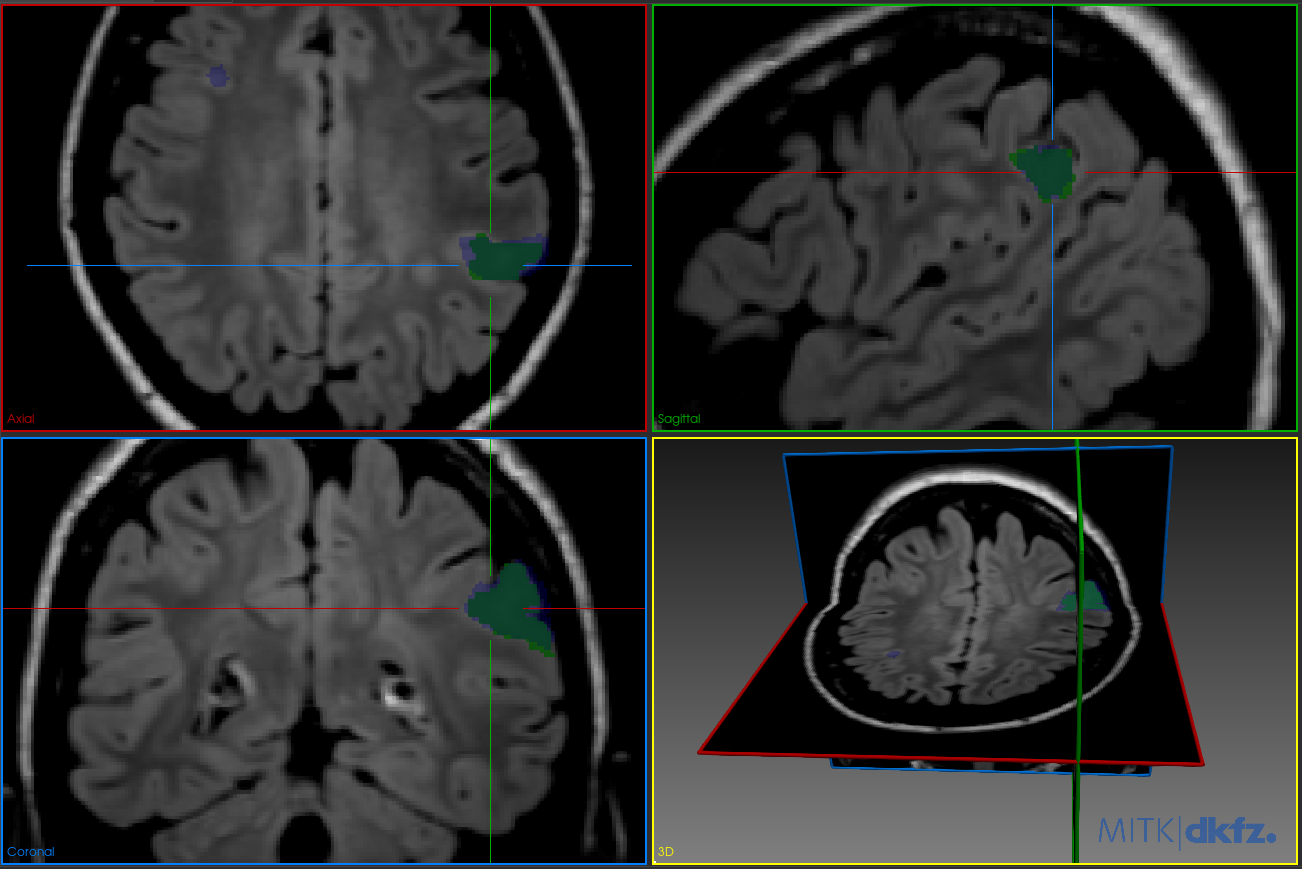}
        \caption{Results of the base model. Green: ground truth mask; blue: predicted mask. Note the false positive cluster in the axial view (top-right pane).}
        \label{fig:1a}
    \end{subfigure}
    \hfill
    \begin{subfigure}[t]{0.48\linewidth}
        \centering
        \includegraphics[width=\linewidth]{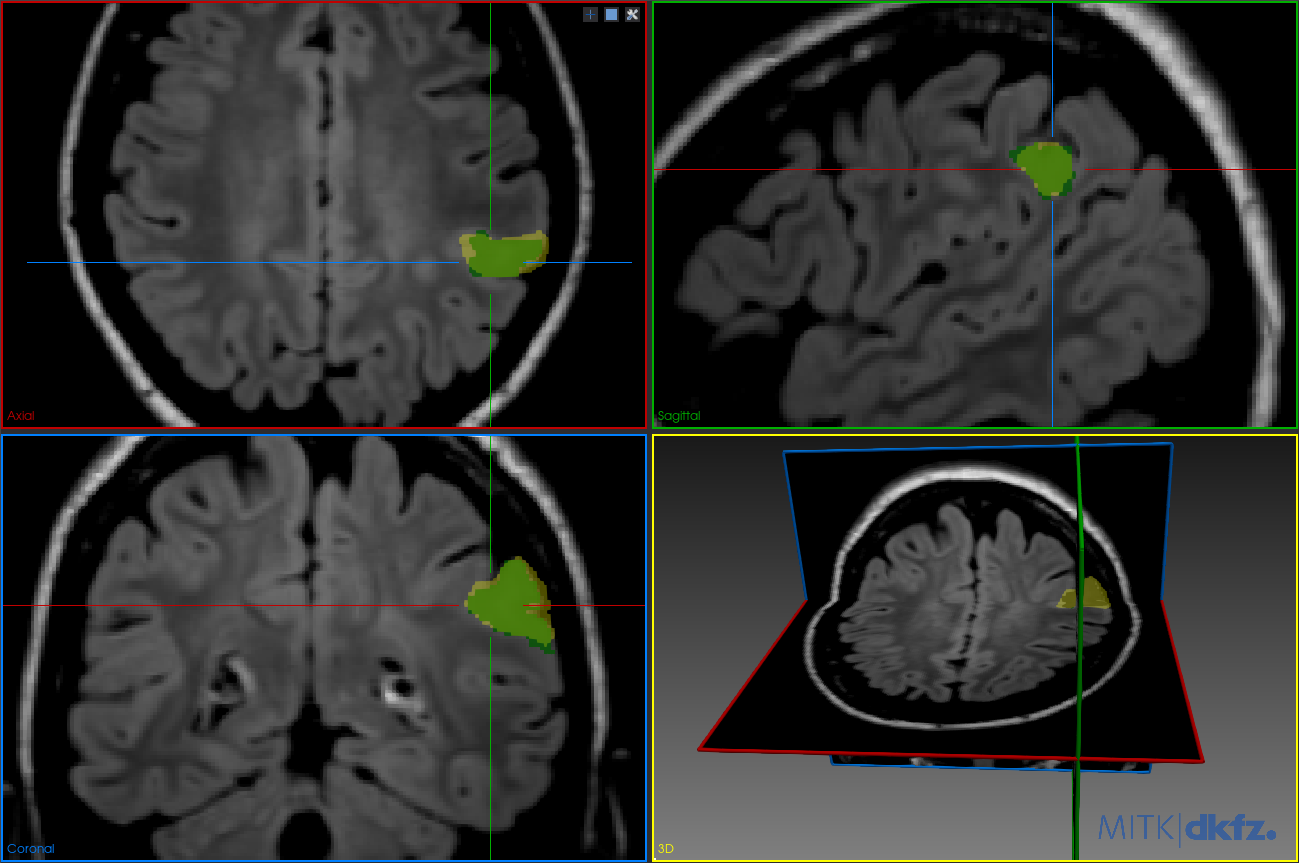}
        \caption{Results after applying post-processing. Green: ground truth mask; yellow: predicted mask. Note that the false detection is removed.}
        \label{fig:1b}
    \end{subfigure}
    \caption{Comparison of predicted segmentation before and after post-processing for the base model.}
    \label{fig:1}
\end{figure}

\begin{figure}[h!]
    \centering
    \begin{subfigure}[t]{0.48\linewidth}
        \centering
        \includegraphics[width=\linewidth]{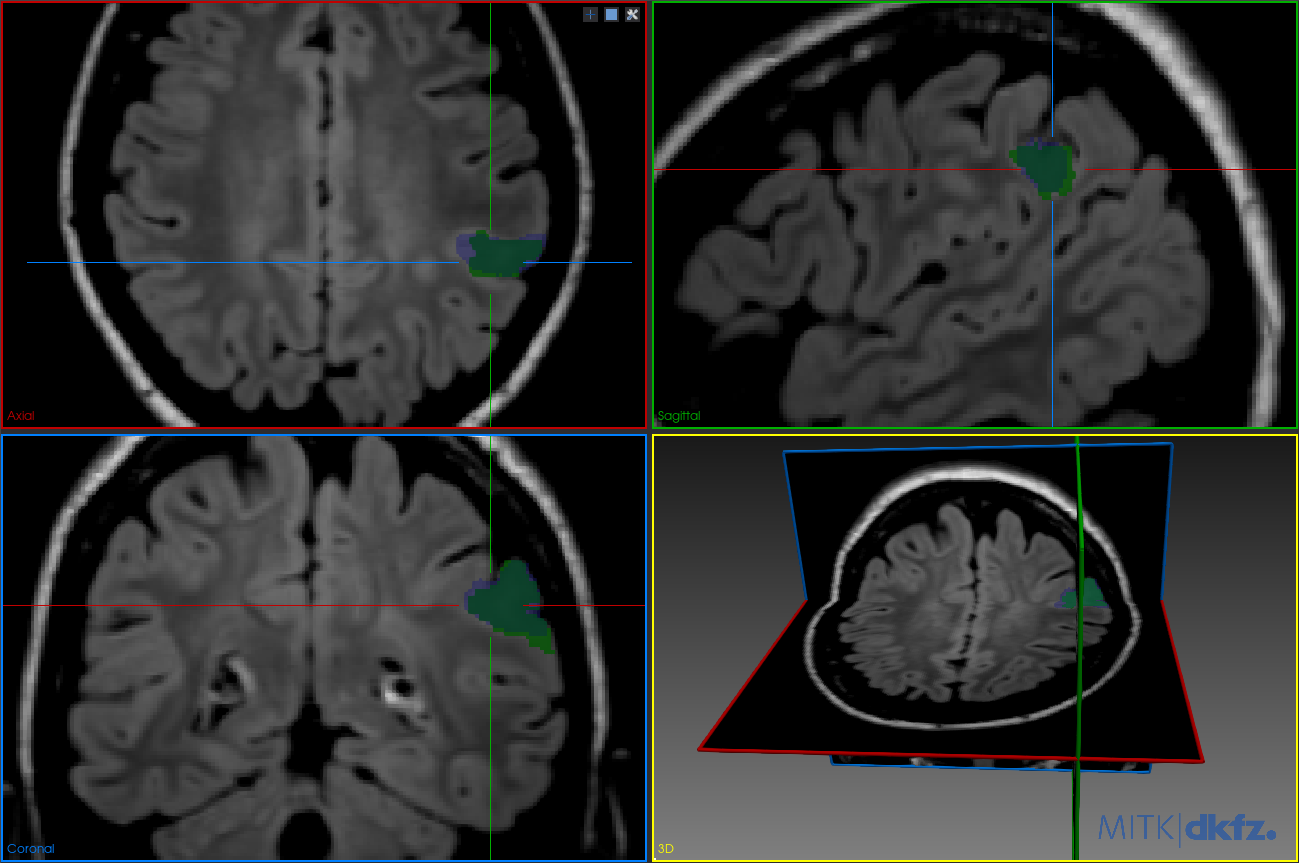}
        \caption{Model trained with TV loss. Green: ground truth mask; blue: predicted mask. The false positives are no longer present.}
        \label{fig:2a}
    \end{subfigure}
    \hfill
    \begin{subfigure}[t]{0.48\linewidth}
        \centering
        \includegraphics[width=\linewidth]{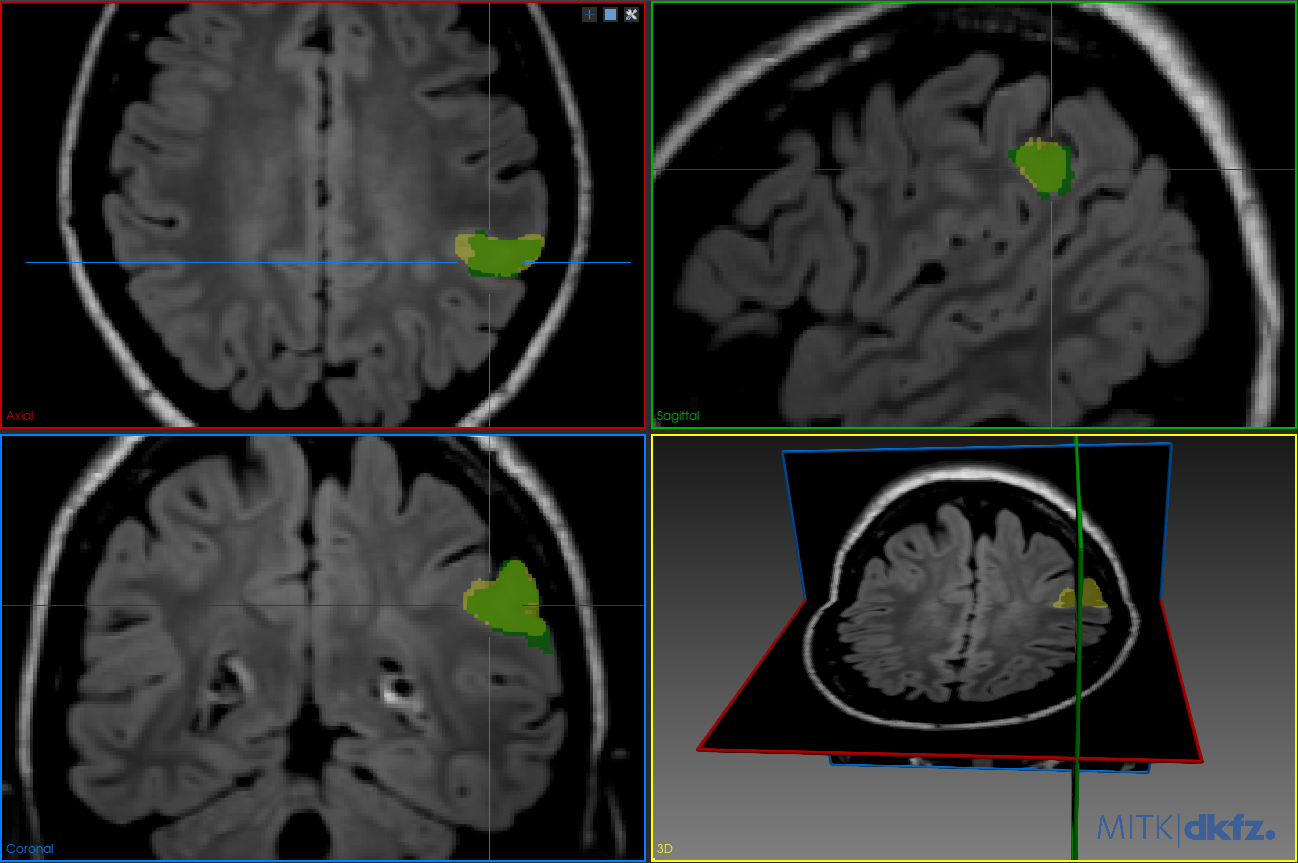}
        \caption{Model trained with TV loss after post-processing. Green: ground truth mask; yellow: predicted mask. Minimal change, indicating TV loss already enforced spatial consistency.}
        \label{fig:2b}
    \end{subfigure}
    \caption{Segmentation results with the proposed TV loss, before and after post-processing.}
    \label{fig:2}
\end{figure}

We visualize the segmentation results on a representative test subject to illustrate the effect of post-processing and the contribution of including TV loss during training (Figure \ref{fig:1} and Figure \ref{fig:2}). Visualizations were generated using the MITK software\footnote{https://github.com/MITK/MITK}.
Figure \ref{fig:1} illustrates the segmentation results produced by the base model trained only with Dice Loss, before and after applying post-processing.
In Figure \ref{fig:1a}, we can see a false positive cluster in the predicted mask (blue) that does not overlap with the ground truth (green), indicating a lack of spatial consistency in the raw network output.
In Figure \ref{fig:1b}, after post-processing, the false positive cluster is successfully removed (yellow mask), confirming that connected component analysis can enforce smoother predictions as a post hoc fix.

Figure \ref{fig:2} illustrates the segmentation results when the model is trained with the proposed TV loss added to Dice loss.
In Figure \ref{fig:2a}, the predicted mask (blue) shows a high degree of overlap with the ground truth (green) and no visible false positives, even without post-processing. This highlights the regularizing effect of TV loss in enforcing spatial smoothness during training.
In Figure \ref{fig:2b}, the result after post-processing (yellow mask) is nearly identical to the unprocessed prediction, confirming that TV loss had already smoothed the prediction to the extent that additional post-processing has minimal impact.

\subsection{Discussion}
Spatial consistency and smoothness in the prediction maps of a volumetric medical image segmentation network are desirable features that can be achieved by applying connected component analysis as a post-processing step on the results or can be seen as a constraint that can guide the network to learn features in a way that creates consistent and smooth outputs.
Comparing the results on a base network that has the best results on the FCD detection task on MRI images in our experiments and adding the TV loss during the training process showed that its improvement to the test metrics is better than applying post-processing. 
%And applying post-processing on a network trained with TV loss doesn’t have that much of an 
{Furthermore, applying post-processing to a model already trained with TV loss yields only a minimal additional improvement effect compared to the base one.} %Therefore, we can say that the smoothness has been improved in the training process and can’t be enhanced so much more.
{Therefore, since the smoothness constraint has already been effectively incorporated during training, leaving little room for further enhancement through post-processing.}

Dice Loss focuses mostly on the intersection between prediction and ground truth mask, while BCE loss encourages the voxel values to be close to 0 or 1 because the ground truth labels are either 0 or 1 for each voxel and the ground truth is essentially smooth and consistent, so at the voxel level BCE loss can help remove small false positive clusters and smooth predictions to an extent, but using TV loss encourages the network more to have smooth transitions of predictions between adjacent voxels.
However, using TV loss could encounter a drawback because it can encourage the network to create an all-zero or all-one output map. This is a trivial solution that has TV loss = 0, and it should be handled by proper weighting of TV loss when it sums up to the original loss. 
Another drawback could be the removal of potentially small true positive regions, because, especially in FCD segmentation, having very small positive regions can be a case.
{It is also worth noting that identifying patients who are harder to treat is a common problem in medical research. Just as some epilepsy patients with FCD are difficult to diagnose and manage, other conditions—such as cardiac patients with allergies to donor organs—face similar challenges. These issues highlight the importance of improving segmentation methods for use in more complex clinical cases~\cite{haynatzki2024building}.
}

\section{Conclusions}\label{sec:5}
This paper introduced a Total Variation (TV) regularized framework for segmenting Focal Cortical Dysplasia (FCD) in three-dimensional (3D) brain MRI data.  Our objective was to tackle a key issue in volumetric medical image segmentation: guaranteeing spatial consistency and anatomical plausibility in the anticipated results.  We incorporated a smoothness requirement into the training process by enhancing a state-of-the-art transformer architecture (MS-DSA-Net) with an anisotropic TV loss term.
Our experimental findings indicate that our straightforward yet efficient regularization approach surpasses conventional post-processing techniques, improving both voxel-level precision and overall segmentation consistency. Notably, we found that models trained using TV loss demonstrated remarkable internal consistency, rendering extra post-processing mostly superfluous—underscoring the efficacy of learning-based regularization.
The proposed method enhances the current initiative to develop resilient and interpretable deep learning systems for clinical neuroimaging applications. 
{Although our method is designed for detecting FCD in brain MRI images, the idea of adding Total Variation regularization can also be useful in other medical imaging tasks. For example, similar challenges exist in detecting small lung nodules or subtle cardiac scars in MRI scans. These tasks also require smooth and spatially coherent segmentations, which our approach supports. Future research can explore how this method performs in such diverse medical imaging problems.}
This methodology establishes a basis for further investigation of integrated regularization techniques, particularly in scenarios with limited training data and reduced lesion visibility.  Future research may explore adaptive or region-specific regularization, incorporation of uncertainty estimation, or extensive evaluation on larger datasets.

\begin{credits}
\subsubsection{\ackname} We acknowledge the support of the PNRR project FAIR - Future AI Research (PE00000013), Spoke 9 - Green-aware AI, under the NRRP MUR program funded by the NextGenerationEU. 
\subsubsection{\discintname}
The authors have no competing interests to declare that are
relevant to the content of this article.
\end{credits}
 \bibliographystyle{splncs04}
 \bibliography{ref}

\end{document}